\documentclass[lettersize,journal]{IEEEtran}
\usepackage{amsmath,amsfonts}
\usepackage{algorithmic}
\usepackage{algorithm}
\usepackage{array}
\usepackage[caption=false,font=normalsize,labelfont=sf,textfont=sf]{subfig}
\usepackage{textcomp}
\usepackage{stfloats}
\usepackage{url}
\usepackage{verbatim}
\usepackage{graphicx}
\usepackage{cite}
\usepackage{multirow}
\begin{document}

\title{A Multi-Characteristic Learning Method with Micro-Doppler Signatures for Pedestrian Identification}

\author{Yu Xiang,~\IEEEmembership{Member,~IEEE,} Yu Huang, Haodong Xu, Guangbo Zhang, and Wenyong Wang,~\IEEEmembership{Member,~IEEE}
	
\thanks{This paper was produced by the IEEE Publication Technology Group. They are in Piscataway, NJ.}
\thanks{Manuscript received April 19, 2021; revised August 16, 2021.}}

\markboth{Journal of \LaTeX\ Class Files,~Vol.~14, No.~8, August~2021}%
{Shell \MakeLowercase{\textit{et al.}}: A Sgample Article Using IEEEtran.cls for IEEE Journals}


\maketitle

\begin{abstract}
The identification of pedestrians using radar micro-Doppler signatures has become a hot topic in recent years. In this paper, we propose a multi-characteristic learning (MCL) model with clusters to jointly learn discrepant pedestrian micro-Doppler signatures and fuse the knowledge learned from each cluster into final decisions. Time-Doppler spectrogram (TDS) and signal statistical features extracted from FMCW radar, as two categories of micro-Doppler signatures, are used in MCL to learn the micro-motion information inside pedestrians’ free walking patterns. The experimental results show that our model achieves a higher accuracy rate and is more stable for pedestrian identification than other studies, which make our model more practical.
\end{abstract}


\section{Introduction}
\IEEEPARstart{P}{edestrian} identification is becoming increasingly popular within automatic vehicles and the intelligent traffic system (ITS). 
Most pedestrian identification methods are primarily reliant on video systems \cite{8986832}, but traditional video systems will be severely hampered in low-light or adverse weather situations (such as rain or heavy fog) \cite{severino2019pedestrian},\cite{ni2020human}.
By contrast, radar as a non-contact device based on electromagnetic wave, can work in harsh environmental conditions, and can even penetrate opaque objects like walls without invading the monitored personal privacy, which makes the use of radar for pedestrian identification a more attractive method \cite{gurbuz2019radar,garcia2014analysis}.

With the advancement of radar systems and signal processing technology, not only the target distance but 
also information such as target speed and moving direction may be retrieved through radar echo signals\cite{li2019research},\cite{xiang2019rear}. Because the target's micro-motion (such as rotation and vibration) modulates the frequency of the radar echo signal, resulting in sidebands with respect to the target's 
Doppler frequency shift. This frequency modulation effect is called the micro-Doppler (m-D) effect \cite{chen2006micro}. With radar system, each pedestrian will produce unique m-D signatures caused by micro-motion of their swinging arms, legs, and torso, that make the m-D can be utilized to detect and recognize people \cite{nanzer2017review}. 

Research on the m-D signatures of human walking was firstly carried out in 1998\cite{chen2014radar}. 
An ultra-wideband (UWB) impulse-based mono-static radar was used in \cite{chang2010human} to distinguish humans from moving non-human objects. \cite{javier2014application} investigated the classification of diverse human activities using linear predictive coding (LPC) and m-D signatures. In 2015, Kim et al. offered a human detection and activity classification solution based on deep learning \cite{kim2015human}. In \cite{seyfiouglu2018deep}, the authors classified the m-D signatures of 12 different human activities using a three-layer deep convolutional auto-encoder (CAE). The researches described above focused on human detection and activity classification, but did not deal with person identification problems.
 
Human identification has  garnered a lot of attention in recent years, with advances in radar technology and artificial intelligence. In \cite{fioranelli2015personnel}, the authors employed a multi-static radar method to collect m-D signatures in order to distinguish three people. Cao et al. \cite{cao2018radar} used a K-band Doppler radar and proposed the deep convolutional neural network (DCNN) recognition approach. The m-D signatures of 22 people walking on a treadmill were collected using a 25GHz CW radar in \cite{abdulatif2019person}. Lang et al. \cite{lang2020person} used a plain convolutional neural network (CNN) with a multi-scale feature aggregation strategy to identify four walking people. These researches listed above did reasonably well, but people's movements were constrained, that means they only approached or stayed away from the radar, or even walked on a treadmill. 

Researches on pedestrian identification problems using m-D signatures with uncontrolled movements of pedestrians has just begun. Vandersmissen et al. \cite{vandersmissen2018indoor} extracted m-D signatures and used gait features to identify five people walking freely indoors using low-power FMCW radar. \cite{ni2020human} utilized a transfer learned fine-tune pre-trained ResNet-50 to identify 20 subjects. However, only the time-Doppler spectrogram (TDS) were used within these two studies.

The contributions of this paper are as follows: 1) A multi-characteristic learning (MCL) model with clusters is proposed to reach the demand for jointly learn discrepant pedestrian m-D signatures and fuse the knowledge learned from each cluster into final decisions. 2) In the proposed MCL, we use two categories of m-D signatures (TDS and signal statistical features) extracted from FMCW radar, instead of one within previous works. 3) We adopt data sets with free walking patterns to train and validate MCL, that make our model more practical.

The remainder of this paper is structured in the following manner. In section II, we propose the MCL model and study the m-D signatures. Section III compares and analyzes the results of experiments. The paper is concluded in section IV.

\section{System Description}
\begin{figure}[!t]
	\centering
	\includegraphics[width=3.48in]{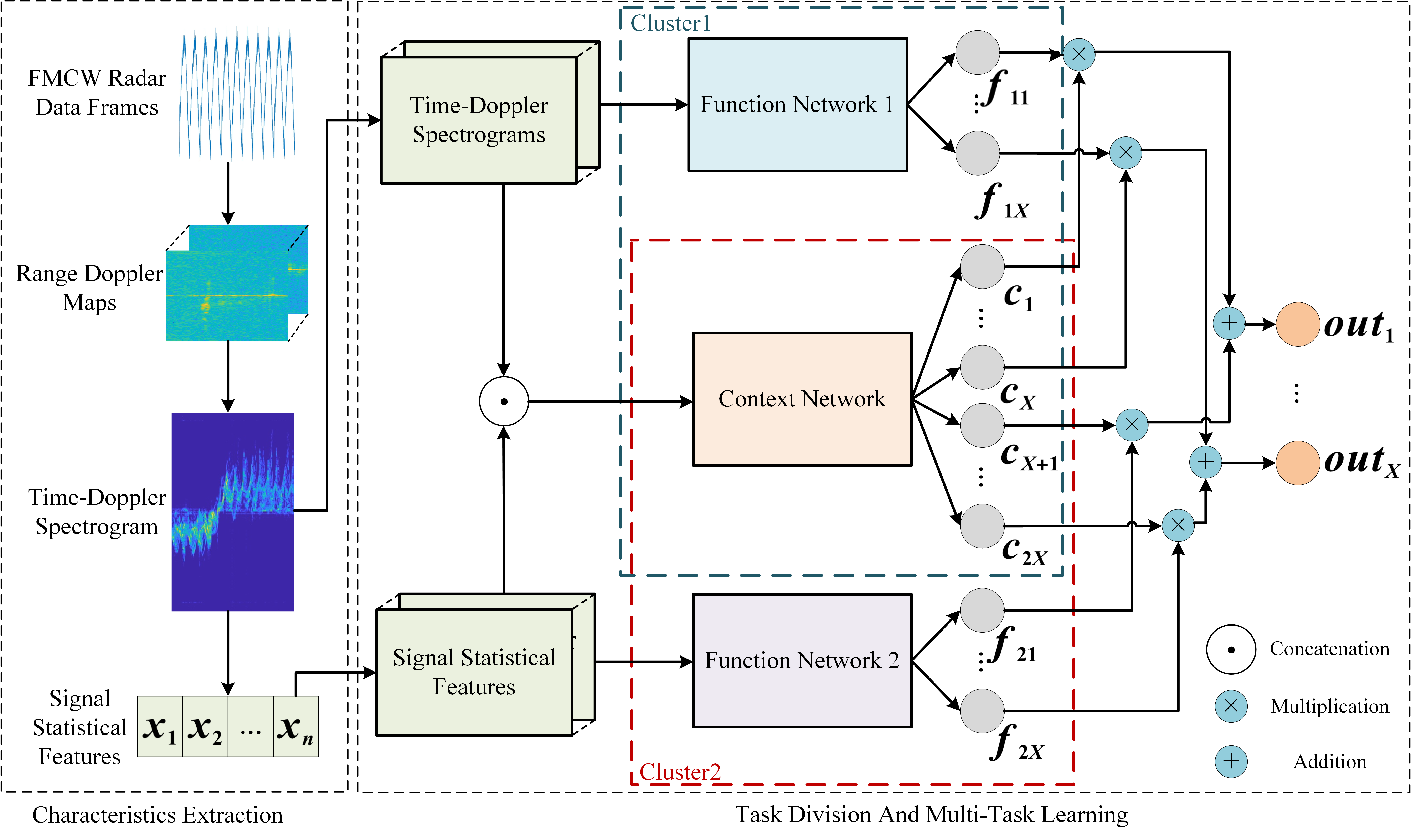}
	\caption{Block Diagram of the MCL.}
	\label{fig_1}
\end{figure}

In this section, we propose a noval deep learning method named multi-characteristic learning (MCL) model for pedestrian identification. MCL model is composed of two modules: one is characteristics extraction (CE) module,  the other is task division and multi-task learning (TD-MTL) module, as shown in Fig.\ref{fig_1}. 
In the CE module, by processing FMCW radar data frames, we acquire two categories of m-D characteristics (i.e., TDS and signal statisticcal features) that we employ as the system's inputs. In TD-MTL module, we construct a multi-characteristic learning network composed of two functional networks (FN) and a context network (CN) to learn two categories of m-D signatures (i.e., characters) we extracted in CE module and their weights to the final pedestrian identification results. Each FN can be integrated with the CN to form a cluster, so the entire learning network we proposed can be divided into two clusters. We allocate these two categories of m-D signatures into two independent subtasks, which are used as the inputs of two clusters. Functional network 1 (FN1) is responsible to recognize pedestrians using the TDS, while functional network 2 (FN2) uses signal statistical features to recognize pedestrians. We also integrate all the m-D signatures as the input to CN and learn the weight that these two m-D signatures' influence on the final decisions. Through this method, we jointly learn the knowledge that is inside the two categories of pedestrian m-D signatures and fuse the results of the two clusters to obtain a more accurate final decision output:
\begin{equation}
	\label{equa_1}
	\left\{\begin{array}{c}
	out_i = f_{1i} * c_i + f_{2i} * c_{X+i} \\ \\
	\mathbf{out} =[out_1,\cdots,out_i,\cdots,out_X], (i = 1, \cdots, X)
	\end{array}\right.
\end{equation}
where $X$ represents the number of output layer nodes both in FN1 and FN2. $f_{1i}$  represents the value of the $i$-th output layer node when the TDS is used to identify pedestrians in FN1, and $f_{2i}$ represents the value of the $i$-th output layer node when the signal statistical features are used to identify pedestrians in FN2. The $i$-th and $X+i$-th weight values learned during joint learning of two m-D signatures in CN are represented by $c_i$ and $c_{X+i}$, respectively. $out_i$ is the $i$-th pedestrian's identification probability, and $\mathbf{out}$ is a vector made up of $X$ pedestrian identification probabilities.

\subsection{Characteristics Extraction Module}
From the original time-varying frequency radar echo signal, TDS and signal statistical features can be extracted that contain a wealth of micro-motion information about the human body.  
\subsubsection{Time-Doppler spectrogram}
\begin{figure}
	\centering
	\includegraphics[width=3.48in]{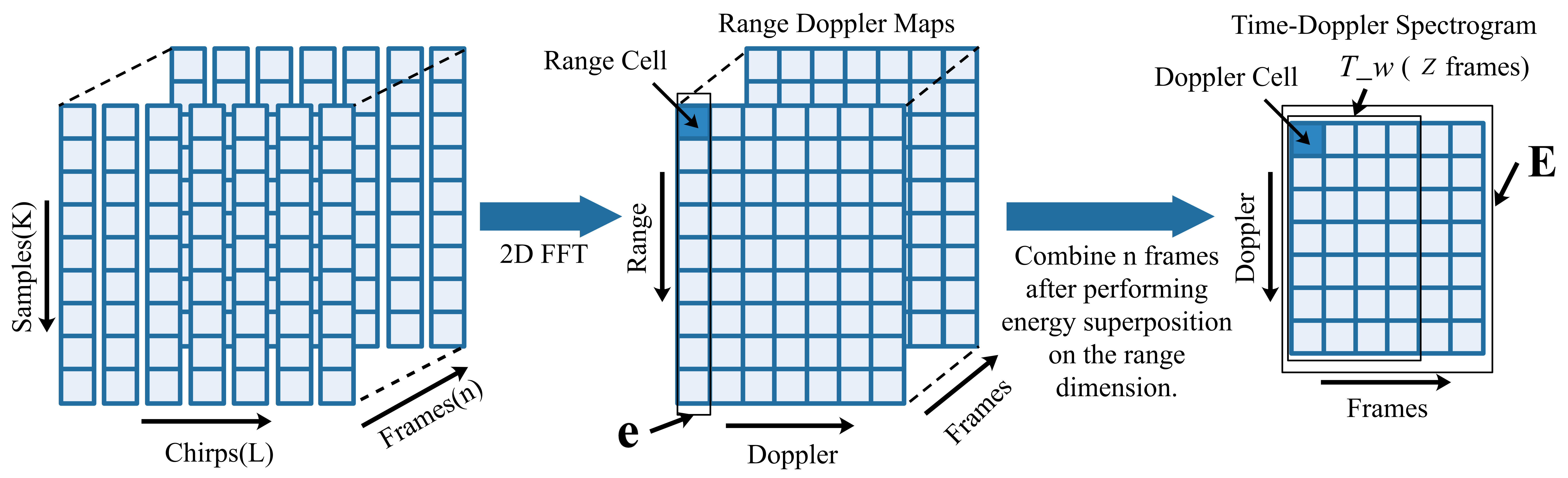}
	\caption{Flow of time-Doppler spectrogram processing.}
	\label{fig_2}
\end{figure}

As illustrated in Fig.\ref{fig_2}, we show the processing flow of how to acquire a TDS. We assume that signal $s(k,l)$ of size $K\times L$ forms a frame, where $K$ donates the number of sampling points and $L$  donates the number of chirps. And the range-doppler map (RDM) $S(u,v)$ can be created using 2D fast fourier transform (2D FFT):
\begin{equation}
\label{equa_4}
S(u,v)=\sum_{l=0}^{L-1} \sum_{k=0}^{K-1} s(k,l) e^{-j 2\pi\left(\frac{uk}{K}+\frac{vl}{L}\right)},(u,v=1,\cdots,K\mid L)
\end{equation}



The TDS is then created by converting the absolute value of all range cells in each frame into decibels, then superimposing all the range cell values in each frame, then combining $n$ time-continuous frames into a TDS, as follows:
\begin{equation}
\label{equa_5}
\left\{\begin{array}{c}
e_v=\sum_{u=1}^{K} 20 \log _{10}|S(u, v)|,(u=1, \cdots, K) \\ \\
\mathbf{e}=[{e_1},\cdots,{e_v},\cdots,{e_L}]^{\mathrm{T}},(v=1, \cdots, L)\\ \\
\mathbf{E}=[\mathbf{e_1}, \mathbf{e_2}, \cdots, \mathbf{e_n}]
\end{array}\right.
\end{equation}
where $e_v$ donates the value of in each doppler cell element, $\mathbf{e}$ donates the energy vector of one frame (i.e., L Doppler cells), $\mathbf{E}$ donates the TDS of $n$ frames, as shown in Fig.\ref{fig_2}.

\begin{figure}
	\centering
	\includegraphics[width=3.48in]{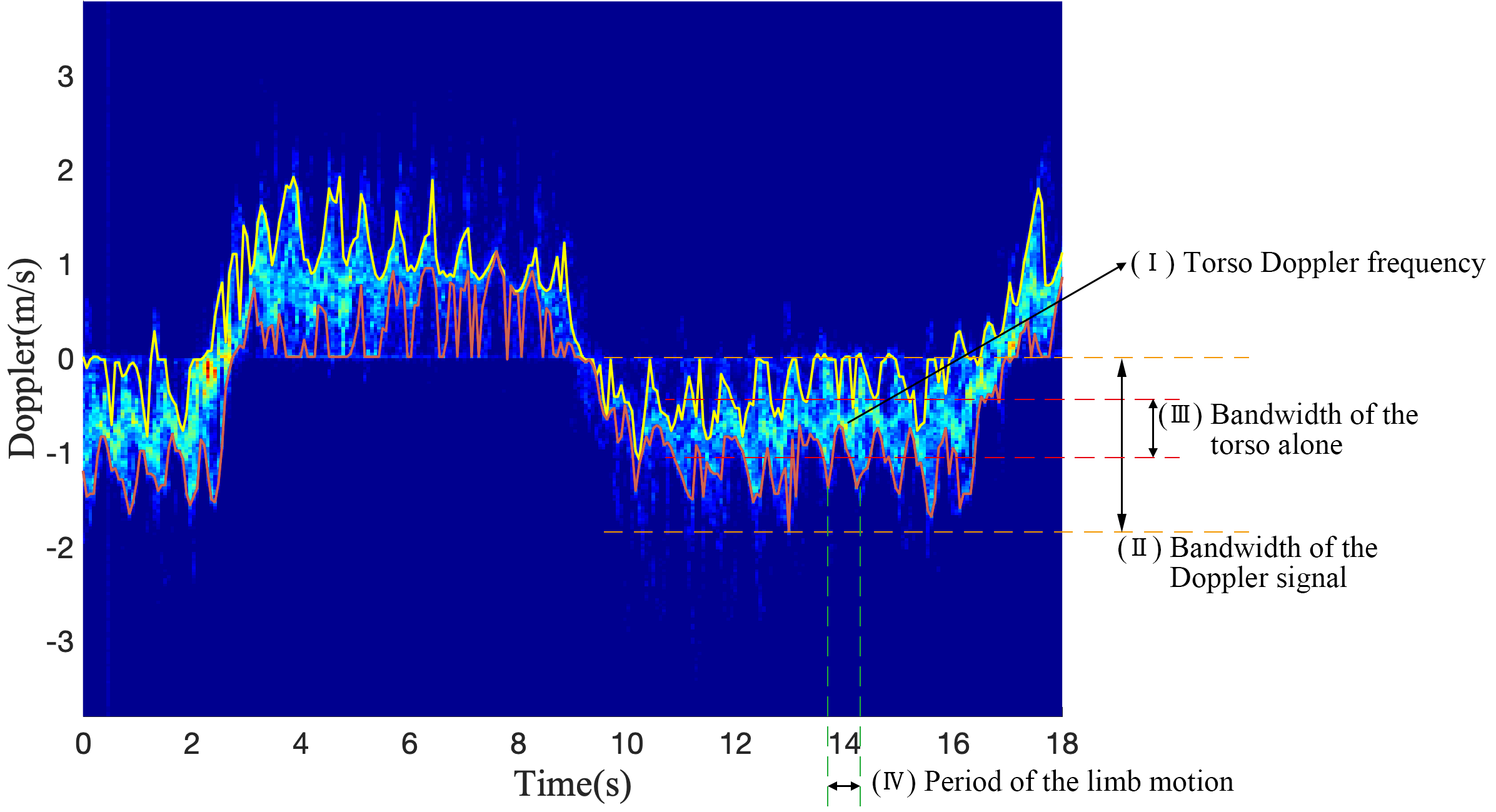}
	\caption{Time-Doppler spectrogram with four features.}
	\label{fig_3}
\end{figure}

Following that, we filtered out the skew-normal distributed noise and deleted the zero Doppler channel representing static objects in the TDS \cite{vandersmissen2018indoor}. In the IDRad data set (explained later in section III), 256 chirps form one frame, with each chirp lasting 256\textmu s. Therefore, 1 second of data contains 15 frames. Fig.\ref{fig_3} depicts an 18-s TDS after previous processing.
\subsubsection{Signal Statistical Features}
To further explore Doppler information to identify the target, we define a time window $T_w$ contains $Z$ time-consecutive frames to calculate four separate signal statistical features from the spectrogram, as follows:

(\uppercase\expandafter{\romannumeral1}) the torso Doppler frequency. It corresponds to a pedestrian's torso speed and can be calculated as (\ref{equa_7}):
\begin{equation}
\label{equa_7}
x_{1}=\frac{\sum_{i=1}^{Z}\left |V_i \right|}{Z}, V_i = \frac{dopplerArgmax(\mathbf{e}_i) \lambda}{2}
\end{equation}
where $\mathbf{e}$ is expressed in (\ref{equa_5}), $dopplerArgmax(\mathbf{e}_i)$ represents the Doppler shift corresponding to the vector $\mathbf{e}$'s maximal signal strength, $V_i$ represents the torso speed in a frame, $\lambda$ represents wavelength. 

(\uppercase\expandafter{\romannumeral2}) the bandwidth of the Doppler signal $x_2$. It corresponds to limb motion speed and can be calculated as (\ref{equa_8}):
\begin{equation}
\label{equa_8}
x_{2}=\max({upperEnv}(\mathbf{E}_{T_w}))-\min({lowerEnv}(\mathbf{E}_{T_w}))
\end{equation}
where $\mathbf{E}_{T_w}$ is a TDS $\mathbf{E}$ within the time window $T_w$.  ${upperEnv}(\mathbf{E}_{T_w})$ and ${lowerEnv}(\mathbf{E}_{T_w})$ represent extracting upper and lower envelopes within time window, respectively.

(\uppercase\expandafter{\romannumeral3}) the bandwidth of the torso alone $x_3$. It corresponds to the Doppler bandwidth without m-D, given by (\ref{equa_9}):
\begin{equation}
\label{equa_9}
x_{3}=\operatorname{avg}({upperEnv}(\mathbf{E}_{T_w}))-\operatorname{avg}({lowerEnv}(\mathbf{E}_{T_w}))
\end{equation}

(\uppercase\expandafter{\romannumeral4}) the period of the limb motion $x_4$. It corresponds to the rate at which the limbs swing, given by (\ref{equa_10}):
\begin{equation}
\label{equa_10}
x_{4}=\frac{T_{w}}{{extremePoint}(\mathbf{E}_{T_w})}
\end{equation}
where $extremePoint$ represents computing the number of extremum point of the upper envelope or the lower envelope.

\subsection{Task Division And Multi-Task Learning (TD-MTL) Module}
Most of the previous research on radar target recognition employed single-task learning (STL) regardless of differences among data characters \cite{lang2019joint}. In our study, TDS and signal statistical features are the two categories of inputs into our proposed model. They are related but depict different aspects of the original signals. Traditional STL is impossible to determine which of the two sets of data has a stronger impact on the final categorization result using its unique output. Multi-task learning (MTL) \cite{ruder2017overview} and cluster networks \cite{white2020fast} have made it possible to learn several related but different tasks at the same time and increase the model's generalization abilities and accuracy. 

\begin{figure}
	\centering
	\includegraphics[width=3.4in]{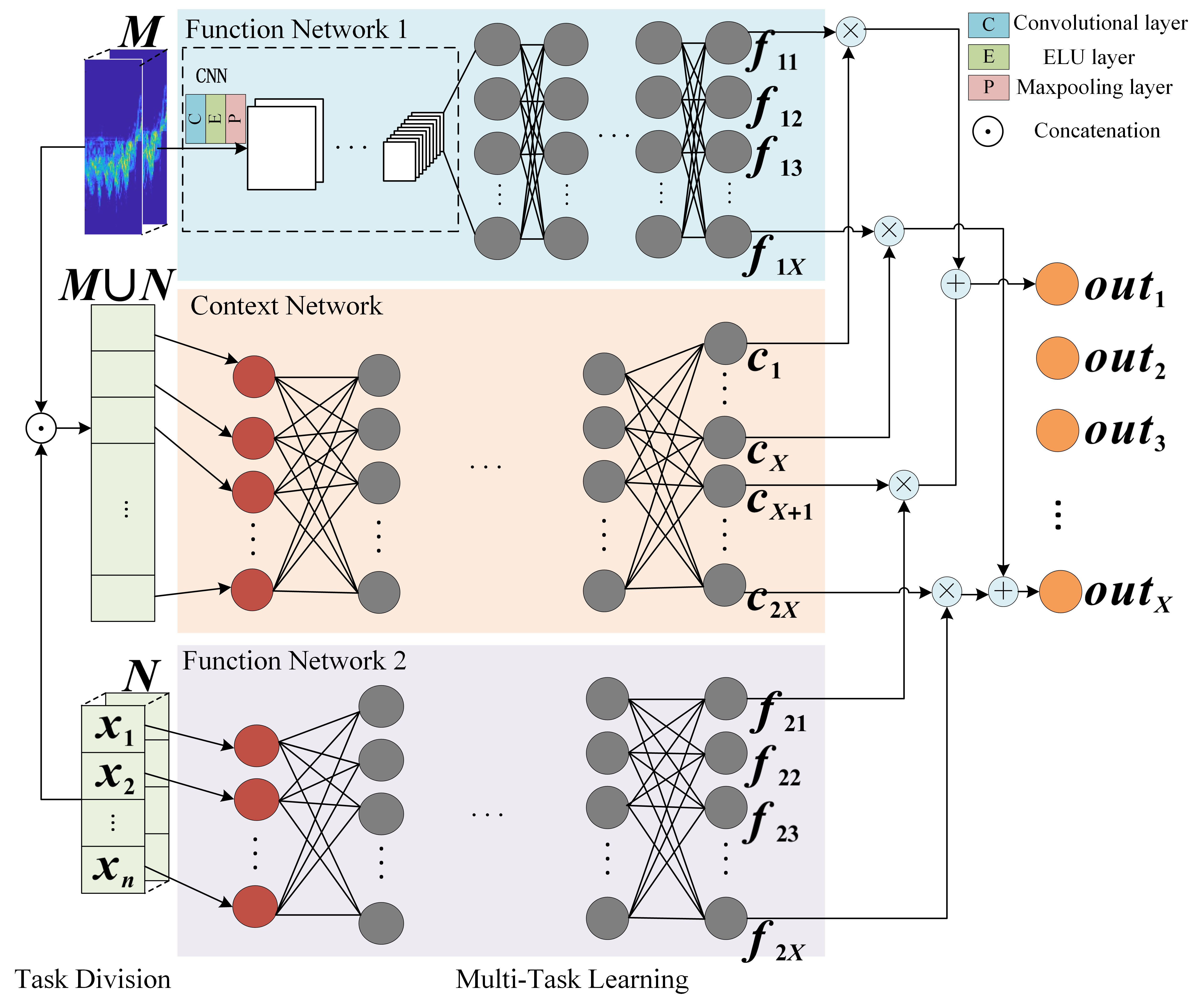}
	\caption{The structure of TD-MTL.}
	\label{fig_4}
\end{figure}

In our proposed model, shown in Fig.\ref{fig_4}, task division (TD) and MTL are the two submodules in TD-MTL.  
\subsubsection{Task Division submodule}
This submodule allocates the two categories of input characters into two subtasks, one for each feed into the two 
functional networks, respectively. The two subtasks used in our model are TDS $M$ 
and signal statistical features $N$, described in the previous section, as follows:
\begin{equation}
\label{equa_11}
M=\bigcup_{i=1}^{P} ({\mathbf{E}_{T_w}})_{i}, N=\bigcup_{i=1}^{P} [x_1,x_2,\cdots,x_n]_i
\end{equation}
where $[x_1,x_2,\cdots,x_n]$ represents $n$ signal statistical features in a $T_w$. The sample numbers of the two subtasks are both represented by $P$.

\subsubsection{Multi-Task Learning submodule}
This submodule is mainly composed of three parts: two functional networks and a context network. In our model, 
we set FN1 as a DCNN, responsible for processing the spectrogram; FN2 as a fully connected network (FCN), responsible for processing statistical features; and CN is also set as a FCN, responsible for learning $M$ and $N$ differentiation and correlations in pedestrian recognition:
\begin{equation}
\label{equa_13}
\left\{\begin{array}{c}
	f_{1i}=FN1(M), i=1,2, \cdots, X \\ \\
	f_{2i}=FN2(N), i=1,2, \cdots, X\\ \\
	c_{i}=CN(M \cup N), i=1,2, \cdots, 2X
	\end{array}\right.
\end{equation}
$f_{1i}$, $f_{2i}$, $c_{i}$, and $X$ are expressed in (\ref{equa_1}). Here, we combine $M$ and $N$ and input them into CN to learn the weights of two categories of m-D signatures to obtain more accurate fusion judgment results. 

The TD-MTL module uses the strategy of training FN and CN together. In the network, we use the cross-entropy loss function, given by (\ref{equa_16}):

\begin{equation}
\label{equa_16}
\operatorname{CE}(\mathbf{label},\mathbf{out})=\frac{1}{P} \sum_{i=1}^{P}\left[-\sum_{j=1}^{Q} {label}_{j|i}\left(\log \left({out}_{j|i}\right)\right)\right]
\end{equation}
where $P$ is the number of samples, $Q$ is the number of categories, ${out}_{j|i}$ is the probability that the $i$-th sample in the TD-MTL will be classified into the $j$-th category. ${label}_{j|i}$is a sign function, if the true class of sample i is equal to j take 1, otherwise take 0.

The complete model training strategy is described in Algorithm 1. In the algorithm, the output layer nodes of FN1 and FN2 are both $X$, $\gamma$ is the learning rate, $\mathbf{W}$ and $\mathbf{b}$ are the weight matrix and bias vector of the entire network, respectively.




\floatname{algorithm}{Algorithm}
\renewcommand{\algorithmicrequire}{\textbf{Initialization:}}
\renewcommand{\algorithmicensure}{\textbf{Input:}}
\begin{algorithm}[H]
	\caption{Training Procedure}
	\begin{algorithmic}[1]
		\REQUIRE $\gamma$, $X$, $\mathbf{W}$, $\mathbf{b}$
		\ENSURE $M$(FN1 Input), $N$(FN2 Input), $M\cup N$(CN Input)
		\FOR{$loss \to 0$}  
			\STATE Forward propagation:     
			\STATE \hspace{0.5cm} FN1 Output: $f_{1i}=FN1(M)$
			\STATE \hspace{0.5cm} FN2 Output: $f_{2i}=FN2(N)$
			\STATE \hspace{0.5cm} CN Output: $c_{i}=CN(M\cup N)$
		
			\STATE \hspace{0.5cm} $\mathbf{out} =[out_1,\cdots,out_X]$
			\STATE Compute $loss=CE(\mathbf{label}, \mathbf{out})$ 
			\STATE Training FN1,FN2 and CN to update parameters:$\mathbf{W}$, $\mathbf{b}$ 
			\STATE \hspace{0.5cm} $\mathbf{W} =\mathbf{W} + \gamma\frac{\partial loss}{\mathbf{W}}$
			\STATE \hspace{0.5cm} $\mathbf{b} =\mathbf{b} + \gamma\frac{\partial loss}{\mathbf{b}}$
		\ENDFOR  
	\end{algorithmic}
\end{algorithm}

\section{Experimental Results And Analysis}
\subsection{Experiment Descriptions}
Our research is based on the open-source data collection IDRad, which contains 150 minutes of tagged pedestrian FMCW radar data, produced by Industrial Radar Systems GmbH. The pedestrian radar data were collected while five people with different body types and similar postures walked freely, that made these data sets more practical. The data sets included a training set of 100 minutes, a validation set of 25 minutes, and a test set of 25 minutes. Each person had 20 minutes of data in training set and 5 minutes of continuous data in the validation and test sets. More details of IDRad can be found in \cite{vandersmissen2018indoor}.

Our proposed MCL structure parameters are summarized in Table \ref{table_2}. CNN $(a, b, c, d)$ represents a four-layer convolutional neural network with number of channels named $a$, $b$, $c$, $d$. FCN$(e,f,g,h)$ represents a fully connected neural network with the number of neurons in each layer from the input layer to the output layer being $e$, $f$, $g$ and $h$. In FN1, the size of the convolution kernel is $3\times3$, after each convolutional layer, there is a maximum pooling layer and an activation layer with an ELU activation function. The TDS input size is $45\times205$, where 45 denotes 3-s data (i.e., 45 frames), and the number of doppler cells is 205. In FN2, four signal statistical features from each of the chosen 165 continuous frames (i.e., $T_w$ = 11-s) are used as input. The starting time of the input of FN1 and FN2 should be synchronized. In CN, its input is a vector that combines the inputs of FN1 and FN2. The learning rate for the entire training procedure is 0.001, with a total of 500 epochs.
\begin{table}[!t]
	\caption{MCL's comprehensive structure. Two functional networks (FN1 and FN2), as well as a context network (CN), make up MCL \label{table_2}}
	\centering
	\resizebox{2in}{!}{
		\begin{tabular}{|c|c|}
			\hline
			Network & Structure\\
			\hline
			\multirow{2}{*}{FN1} & CNN (1,16,32,64)\\
			\cline{2-2}
				 & FCN (1600,128,5)\\
			\hline
			FN2 & FCN (4,5,5,5)\\
			\hline
			CN & FCN (9229,1000,100,10)\\
			\hline
		\end{tabular}
	}
\end{table}

\subsection{Results And Analysis}
\begin{figure}[!h]
	\subfloat[\scriptsize\normalfont Accuracy]{\includegraphics[width=0.24\textwidth]{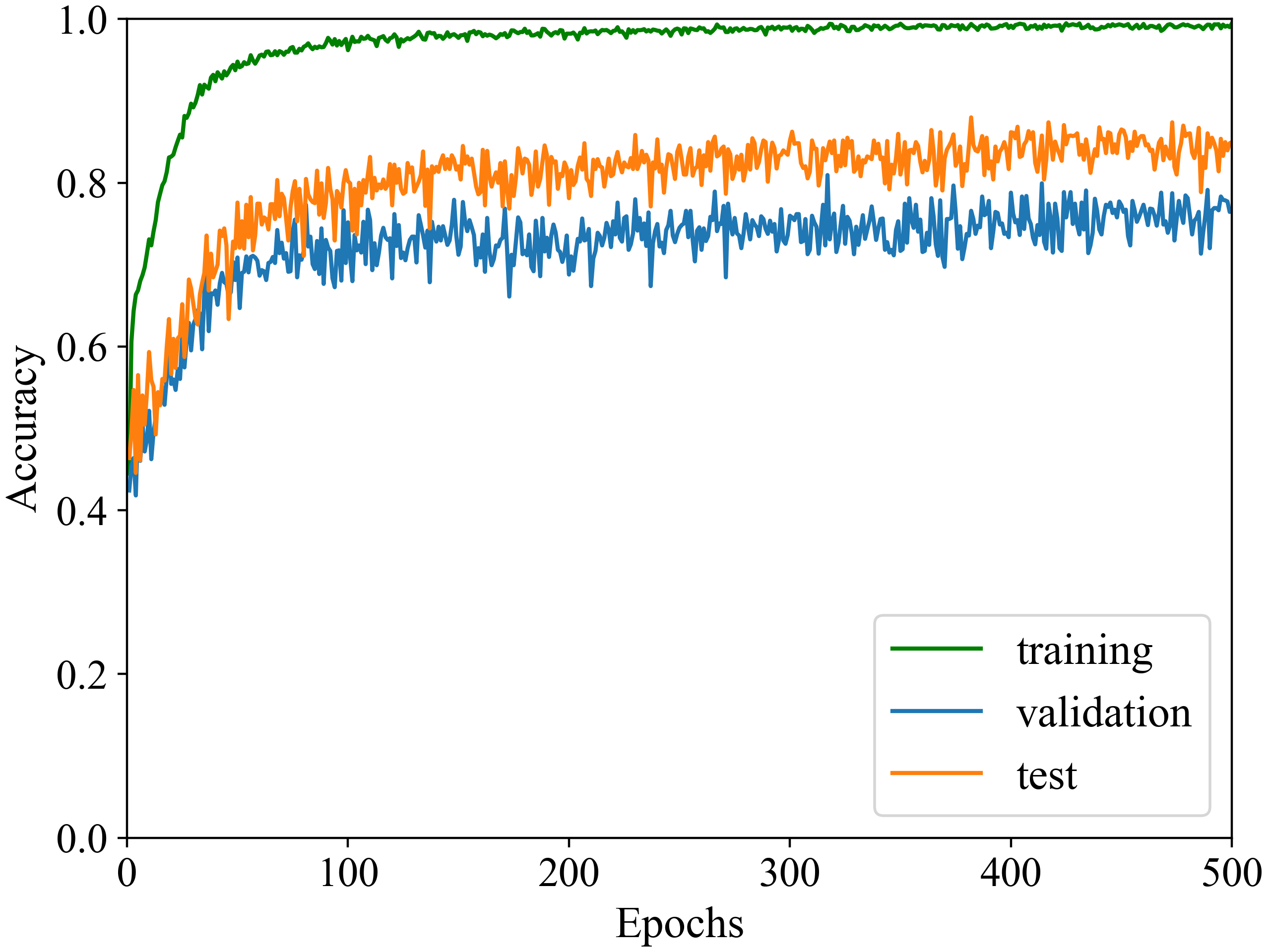}}
	\hfil
	\subfloat[\scriptsize\normalfont Loss]{\includegraphics[width=0.24\textwidth]{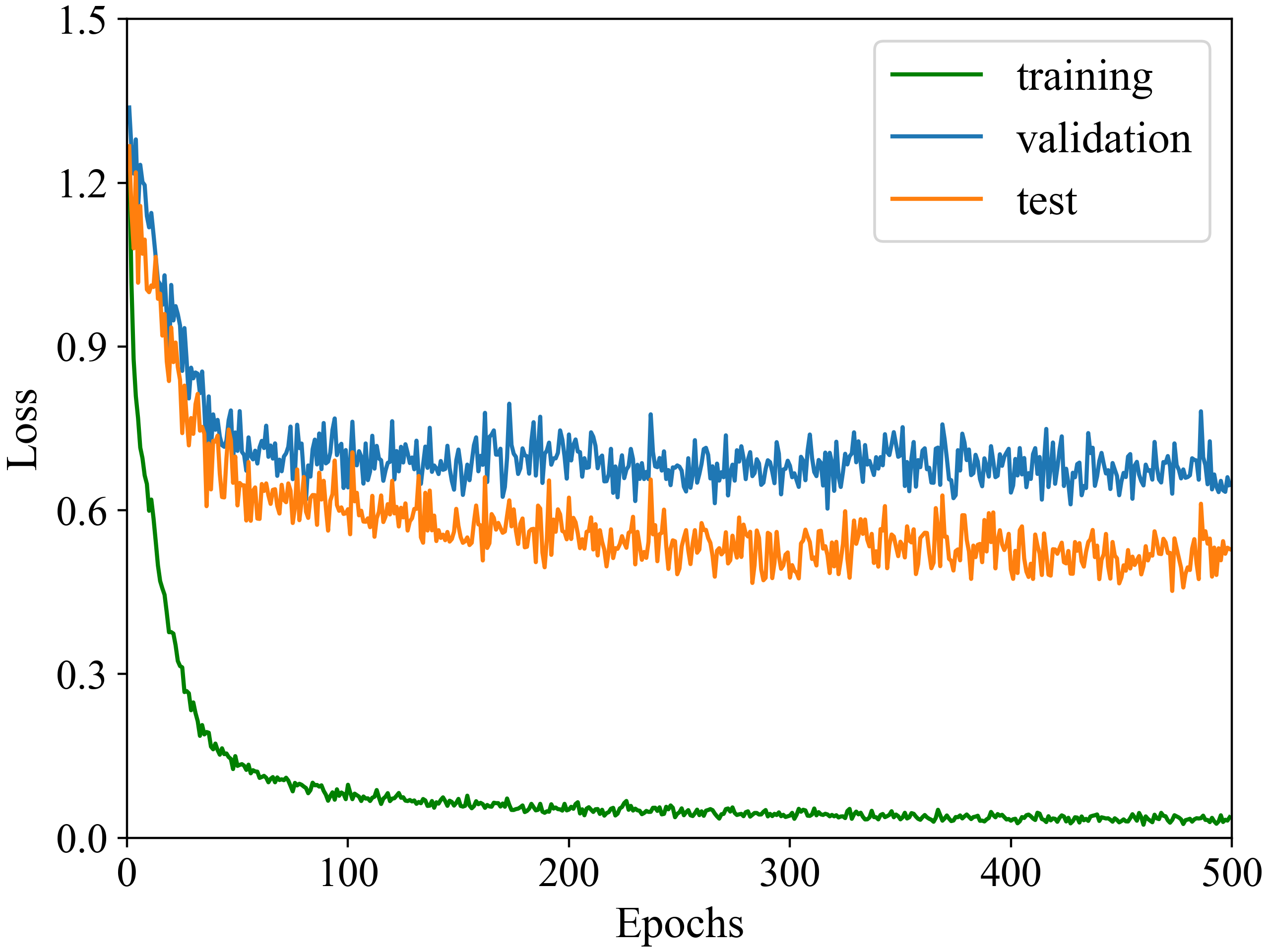}} 
	\caption{Accuracy and loss curves of epochs of training.}
	\label{fig_5}
\end{figure}

Our proposed model obtains an accuracy rate of 87.63\% for identified five pedestrians with the test set and  80.78\% with the validation set, as shown in Fig.\ref{fig_5} (a). The loss curves of the training set, validation set, and test set along with epochs are shown in Fig.\ref{fig_5} (b). We can see that our model converged after $100$-th epoch.

\begin{figure}[!h]
	\subfloat[\scriptsize\normalfont MCL]{\includegraphics[width=0.24\textwidth]{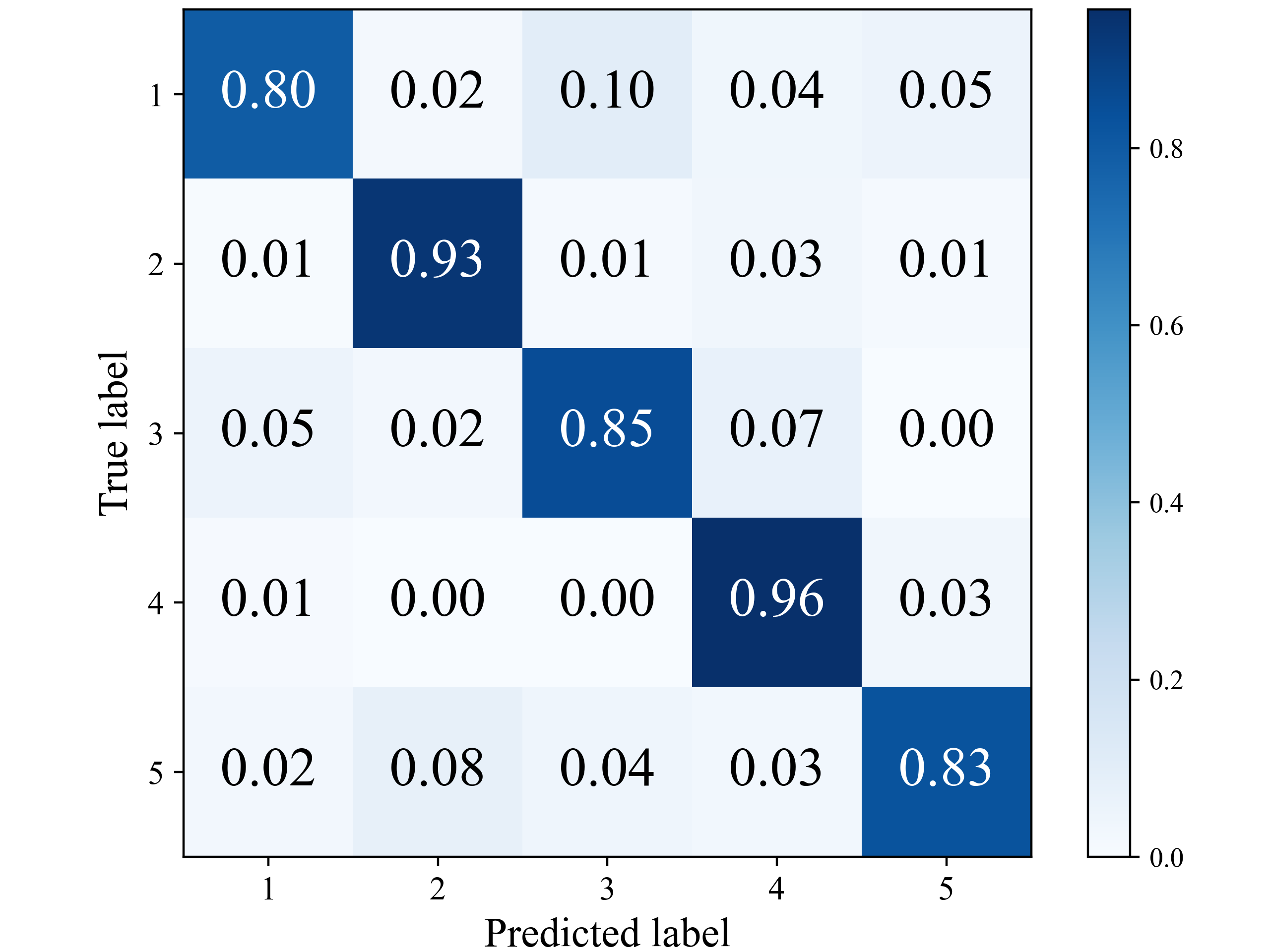}}
	\hfil
	\subfloat[\scriptsize\normalfont Vandersmissen et al.]{\includegraphics[width=0.24\textwidth]{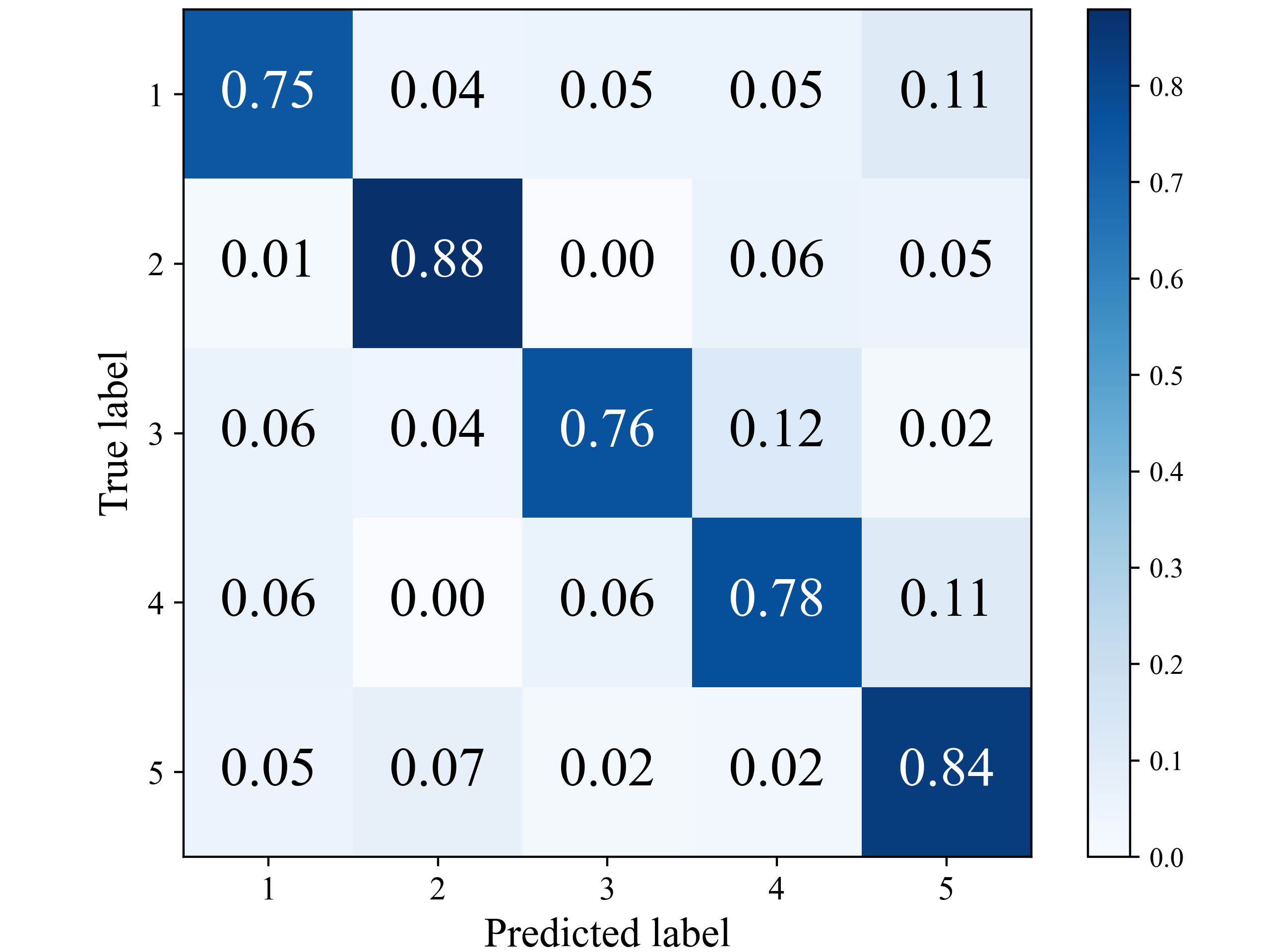}} 
	\caption{Confusion matrix comparison for five pedestrians in test set.}
	\label{fig_6}
\end{figure}

Confusion matrices are shown in Fig.\ref{fig_6}. (a) shows the results from the identification of five pedestrians with the MCL model, (b) shows the results from the identification of five pedestrians in \cite{vandersmissen2018indoor}. As can be observed in (a), our model has a relatively high identification rate for five pedestrians, especially for pedestrians 2 and 4, where the recognition accuracy rate can exceed 93\%. The identification accuracy of pedestrians 1, 3, and 5 is relatively low, as 80\%, 85\%, and 83\%, respectively. These may be due to the pedestrians 1, 3, and 5 are of similar age, height, and weight. Compared to [19], the accuracy rate of pedestrian identification in 3 and 4 has increased by about 9\% and 18\%, respectively. The rise was likewise around 5\% for pedestrians 1 and 2. We can also conclude from Fig.6 that the false identification rates among pedestrians decreased from 1\% to 8\%.

Next, we compared our proposed MCL model with other methodologies commonly used in the field of pedestrian identification. Comparison methods are summarized in Table \ref{table_3}.
\begin{table}[!th]
	\caption{ Comparison Methods \label{table_3}}
	\centering
	\resizebox{3in}{!}{
		\begin{tabular}{|c|c|}
			\hline
			Comparison Method & Network Description\\
			\hline
			\multirow{2}{*}{Vandersmissen et al. \cite{vandersmissen2018indoor} } &  A 6-layer CNN which has four convolutional \\ & layers and two  fully connected layers \\
			\hline
			\multirow{2}{*}{Cao et al. \cite{cao2018radar}} & AlexNet, which has five convolution  \\ & layers  and two fully connected layers\\
			\hline
			\multirow{2}{*}{Lang et al. \cite{lang2020person}} & A plain convolutional neural network with \\ &
			a multi-scale feature aggregation strategy\\
			\hline
			Abdulatif et al. \cite{abdulatif2019person} & ResNet-50: a 50-layer deep residual network\\
			\hline
		\end{tabular}
	}
\end{table}

\begin{figure}[!]
	\centering
	\includegraphics[width=2.3in]{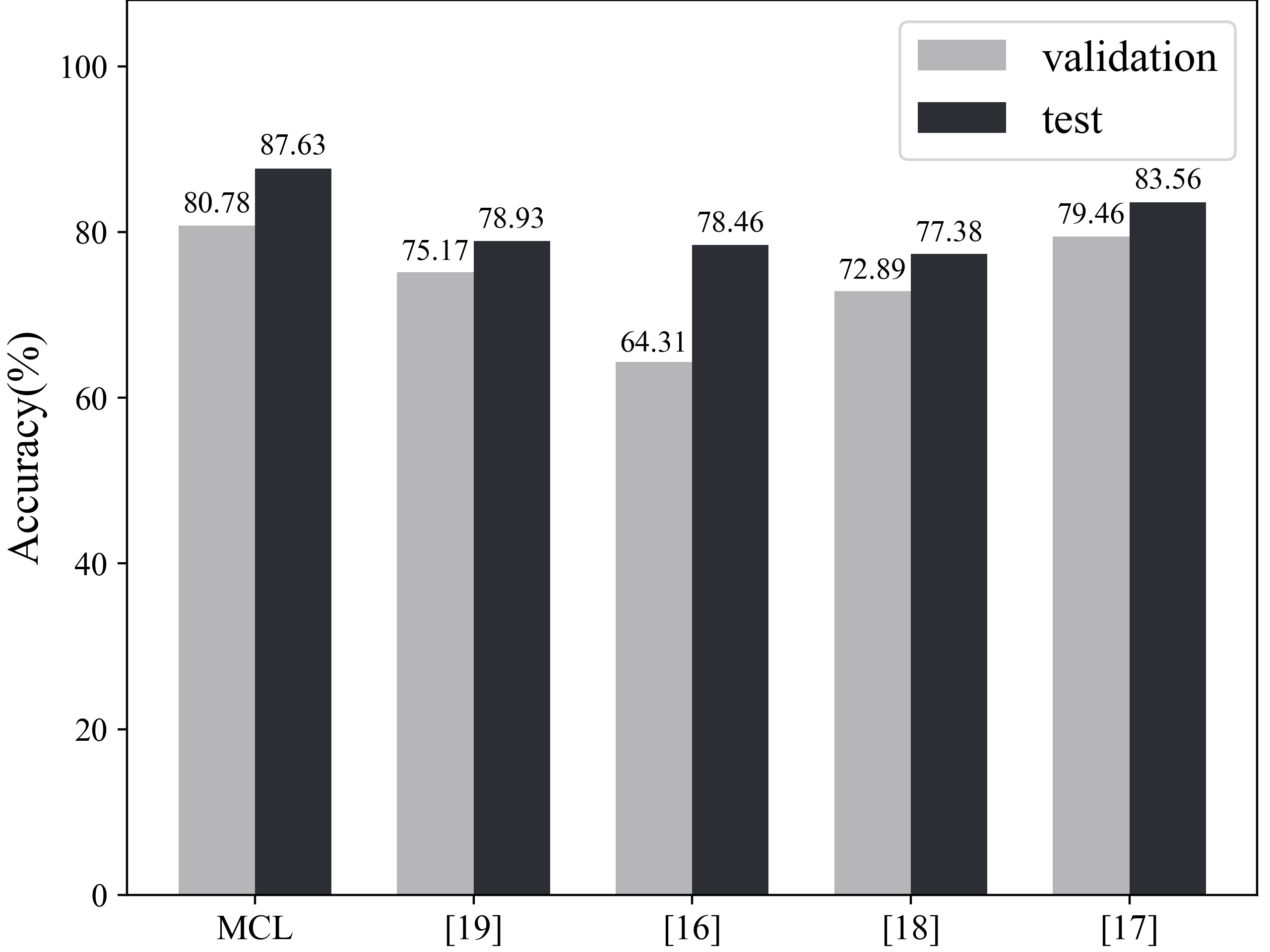}
	\caption{Accuracy rate comparison for pedestrian identification.}
	\label{fig_7}
\end{figure}

The accuracy of our proposed MCL model is superior to various other methods in both validation and test sets used in pedestrian recognition, as shown in Fig.\ref{fig_7}. With validation set, the performance of our model improved from 1\% to 15\%, while with test set, the performance improved from 4\% to 10\%.

\begin{figure}[!]
	\centering
	\includegraphics[width=2.3in]{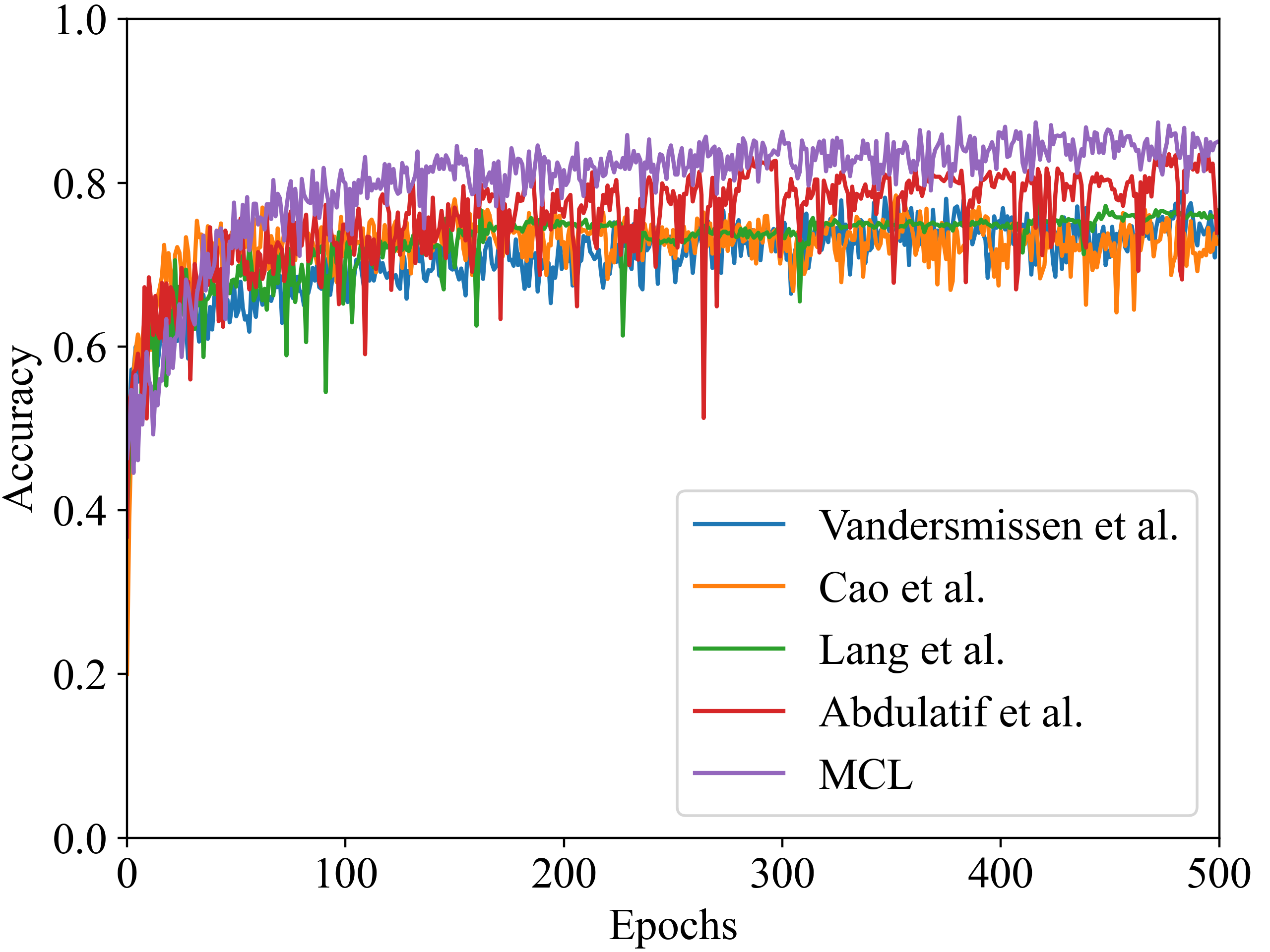}
	\caption{Pedestrian identification accuracy rate comparisons over epochs.}
	\label{fig_8}
\end{figure}

Fig.\ref{fig_8} shows how the test accuracy of our model and each of the comparison methods changed over epochs. Our proposed model achieves the highest model accuracy rate of 87.63\%. Our model also exceeds previous recognition algorithms in terms of stability.
\section{Conclusions}
In this paper, we propose a learning method that combines the ideas of multi-task learning and cluster network to deal with the problem of pedestrian identification walk freely using two categories of m-D signatures. Experimental results show that our proposed method is superior to other recognition methods in terms of accuracy and stability. Our model improved by 1\%-15\% on the validation set, while on the test set, it improved by 4\%-10\%. As a result, the Multi-Characteristic Learning model we suggested can be employed in the future for pedestrian identification.


 

 \begin{IEEEbiography}[{\includegraphics[width=1in,height=1.25in,clip,keepaspectratio]{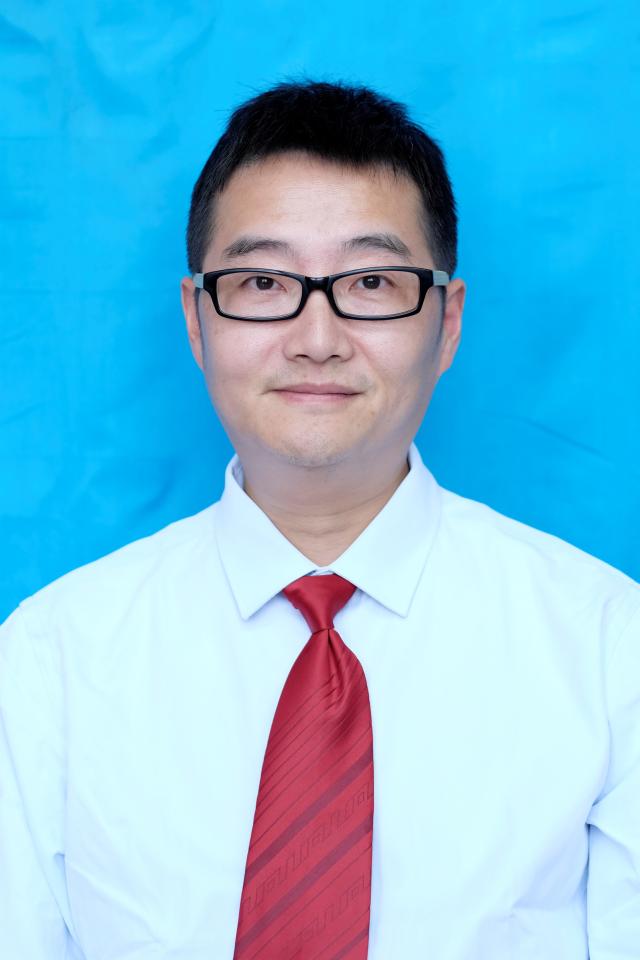}}]{Yu Xiang}
(Member, IEEE) received B.S., M.S., and Ph.D. degrees from the University of Electronic Science and Technology of China (UESTC), Chengdu, Sichuan, China in 1995, 1998, and 2003, respectively. He joined the UESTC in 2003 and became an associate professor in 2006. From 2014 to 2015, he was a visiting scholar with the University of Melbourne, Australia. 

His current research interests include computer networks, intelligent transportation systems and deep learning.
 \end{IEEEbiography}

 \begin{IEEEbiography}[{\includegraphics[width=1in,height=1.25in,clip,keepaspectratio]{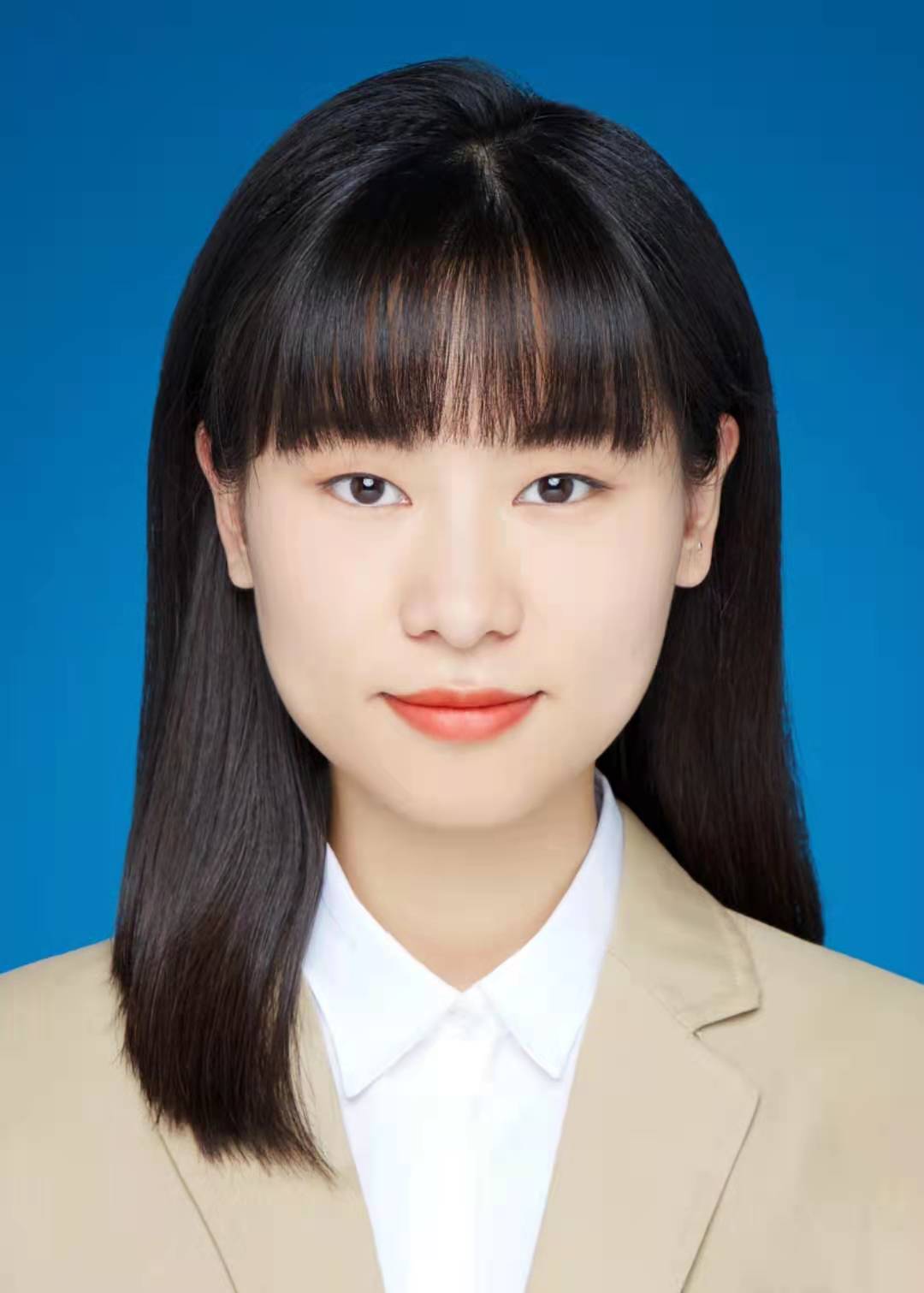}}]{Yu Huang}
received a B.E. degree from the Sichuan Normal University, Chengdu, China in 2019. She is currently pursuing an M.S. degree in computer science at the University of Electronic Science and Technology of China (UESTC), Chengdu, China.

Her current research interests include the Internet of Things and ITS. 
\end{IEEEbiography}

 \begin{IEEEbiography}[{\includegraphics[width=1in,height=1.25in,clip,keepaspectratio]{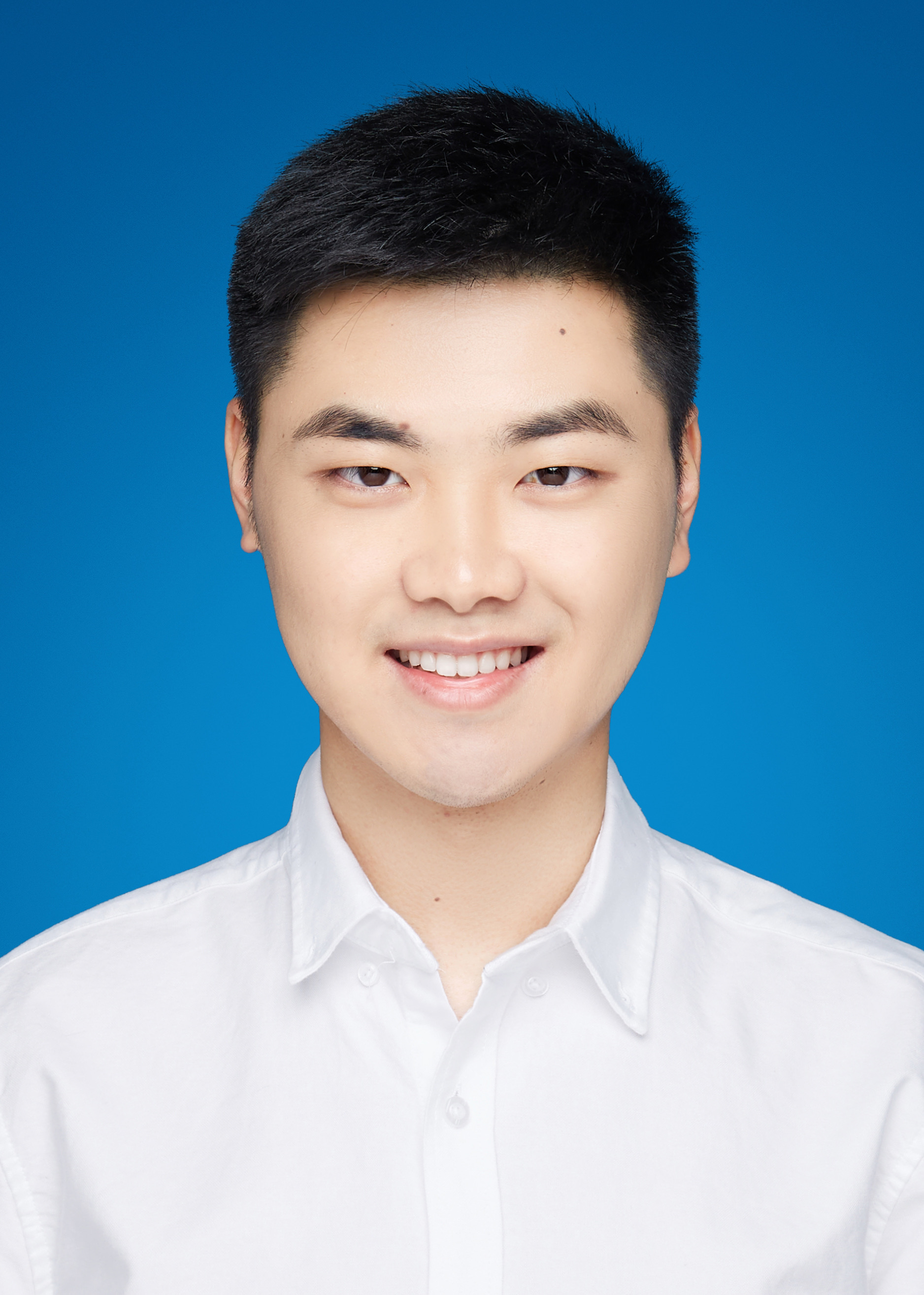}}]{Haodong Xu}
received a B.E. degree from the Xiamen University, Xiamen, China in 2019. He is currently pursuing an M.S. degree in computer science at the University of Electronic Science and Technology of China (UESTC), Chengdu, China.

His current research interests include the Internet of Things and ITS. 

\end{IEEEbiography}

 \begin{IEEEbiography}[{\includegraphics[width=1in,height=1.25in,clip,keepaspectratio]{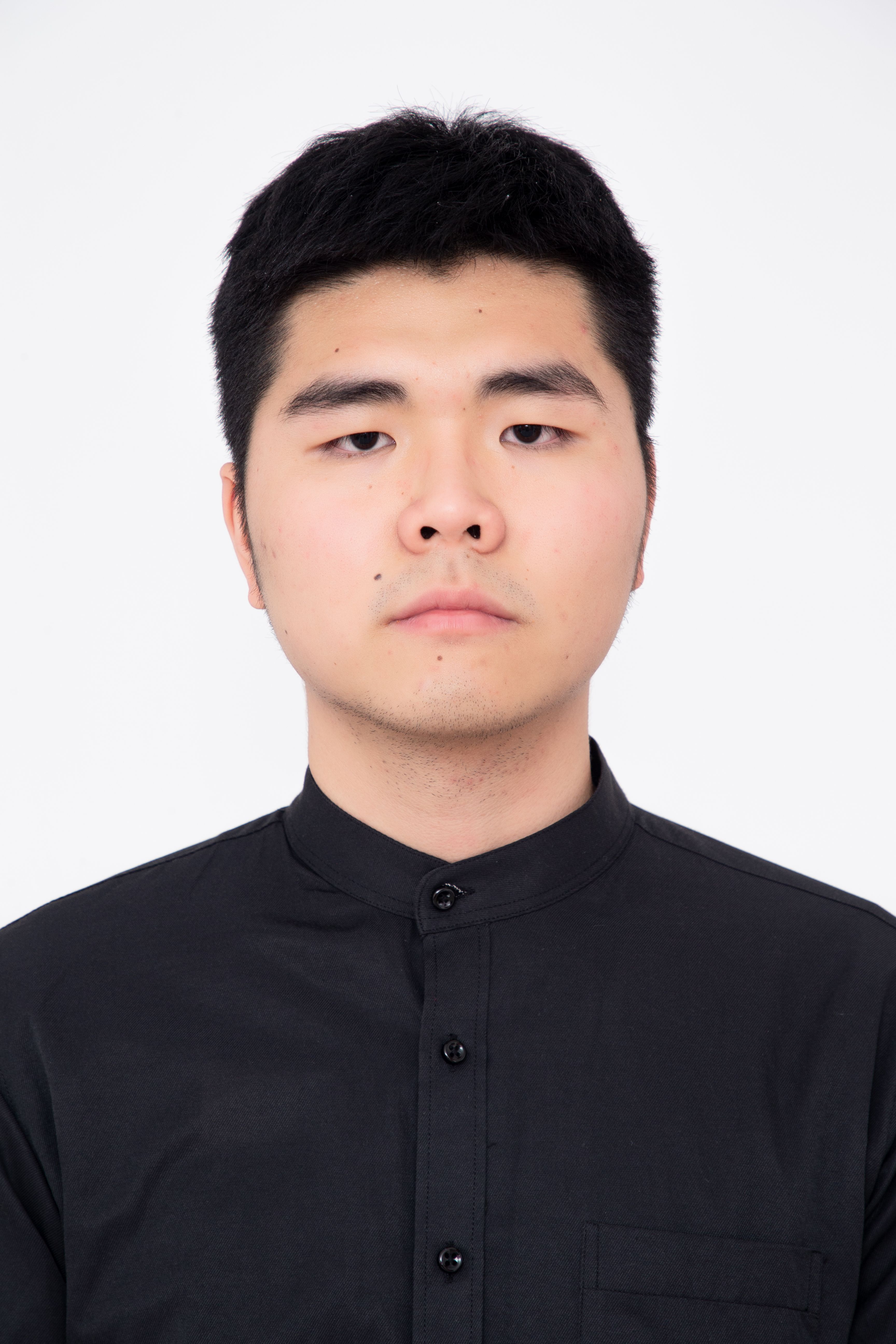}}]{Guangbo Zhang}
received a B.E. degree from the South-Central Minzu University, Wuhan, China in 2020. He is currently pursuing an M.S. degree in computer science at the University of Electronic Science and Technology of China (UESTC), Chengdu, China.

His current research interests include the Internet of Things and deep learning. 
\end{IEEEbiography}

 \begin{IEEEbiography}[{\includegraphics[width=1in,height=1.25in,clip,keepaspectratio]{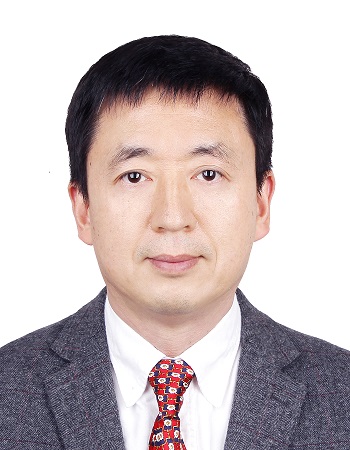}}]{Wenyong Wang}
(Member, IEEE) received a B.S. degree in computer science from Beihang University, Beijing, China in 1988 and M.S. and Ph.D. degrees from the University of Electronic Science and Technology (UESTC), Chengdu, China in 1991 and 2011, respectively. He has been a professor with the School of Computer Science and Engineering, UESTC since 2009. He has served as the director of the Information Center of UESTC and the chairman of the UESTC-Dongguan Information Engineering Research Institute from 2003 to 2009. He is currently a visiting professor with the Macau University of Technology. His main research interests include next-generation Internet, software-designed networks, software engineering, and artificial intelligence. He is a member of the expert board of CERNET and China Next-Generation Internet Committee and a senior member of the Chinese Computer Federation.
\end{IEEEbiography}





\end{document}